\documentstyle[amsmath,epsfig]{aipproc}

\begin{document}
\title{Gamma-Ray Bursts via Pair Plasma Fireballs from Heated Neutron Stars}

\author{Jay D. Salmonson$^*$, James R. Wilson$^{*\dagger}$ and Grant J. Mathews$^{\dagger}$}
\address{$^*$Lawrence Livermore National Laboratory, Livermore, CA 94550\\
$^{\dagger}$University of Notre Dame, Notre Dame, IN 46556}

\maketitle

\begin{abstract}
In this paper we model the emission from a relativistically expanding
$e^+e^-$ pair plasma fireball originating near the surface of a heated
neutron star.  This pair fireball is deposited via the annihilation of
neutrino pairs emanating from the surface of the hot neutron star.
The heating of neutron stars may occur in close neutron star binary
systems near their last stable orbit.  We model the relativistic
expansion and subsequent emission of the plasma and find $\sim 10^{51}
- 10^{52}$ ergs in $\gamma$-rays are produced with spectral and
temporal properties consistent with observed gamma-ray bursts.
\end{abstract}

\section*{Introduction}
It has been speculated for some time that inspiraling neutron stars
could provide a power source for cosmological gamma-ray bursts
\cite{jwg:mr92,jwg:piran98}.  However, previous Newtonian and
post-Newtonian studies \cite{jwg:jr96} of the final merger of two
neutron stars have found that the neutrino emission time scales are so
short that it would be difficult to drive a gamma-ray burst from this
source.  It is clear that a mechanism is required for extending the
duration of energetic neutrino emission.  A number of possibilities
could be envisioned, for example, neutrino emission powered by
accretion shocks, MHD or tidal interactions between the neutron stars,
etc.  The present study, however, has been primarily motivated by
numerical studies of the strong field relativistic hydrodynamics of
close neutron star binaries (NSBs) in three spatial dimensions.  These
studies \cite{jwg:wm95,jwg:wmm96,jwg:mw97,jwg:mmw98a} suggest that
neutron stars in a close binary can experience relativistic
compression and heating over a period of seconds.  During the
compression phase released gravitational binding energy can be
converted into internal energy.  Subsequently, up to $10^{53}$ ergs in
thermally produced neutrinos can be emitted before the stars collapse
\cite{jwg:mw97}.  Here we briefly summarize the physical basis of this
model and numerically explore its consequences for the development of
an $e^+e^-$ plasma and associated GRB.

In \cite{jwg:mw97} properties of equal-mass neutron-star binaries were
computed as a function of mass and EOS (Equation of State).  From
these studies it was deduced that compression, heating and collapse
could occur a few seconds before binary merger.  Our calculation of
the rates of released binding energy and neutron star cooling suggests
that interior temperatures as hot as 70 MeV are achieved.  This leads
to a high neutrino luminosity which peaks at $L_\nu \sim 10^{53}$ ergs
sec$^{-1}$.  This much neutrino luminosity would partially convert to
an $e^+e^-$ pair plasma above the stars as is also observed above the
nascent neutron star in supernova simulations \cite{jwg:wm93}.

\section*{ Neutrino Annihilation and Pair Creation }

Having outlined a mechanism by which neutrino luminosities of
10$^{52}$ to 10$^{53}$ ergs/sec may arise from binary neutron stars
approaching their final orbits, we must calculate the efficiency of
conversion of neutrino pairs into an electron pair plasma via
$\nu\overline{\nu} \rightarrow e^+e^-$.  Here we argue that the
efficiency for converting these neutrinos into pair plasma is probably
quite high.  Neutrinos emerging from the stars will deposit energy
outside the stars predominantly by $\nu\overline{\nu}$ annihilation to
form electron pairs. A secondary mechanism for energy deposition is
the scattering of neutrinos from the $e^+e^-$ pairs.  Strong
gravitational fields near the stars will bend the neutrino
trajectories.  This greatly enhances the annihilation and scattering
rates \cite{jwg:sw99}. For our employed neutron-star equations of
state the radius to mass ratio is typically between $R/M \sim 3$ and 4
just before stellar collapse (in units $G=c=1$).  In \cite{jwg:sw99}
it is shown that $\nu\overline{\nu}$ annihilation rates will be
enhanced by a factor ${\mathcal{F}}(R/M) \sim 8$ to $28$ due to
relativistic effects. From Eq.~24 of \cite{jwg:sw99} we obtain,
\begin{equation}
        \frac{ \dot{Q} }{ L_\nu } \approx 0.03 {\mathcal{F}}(R/M)
L_{53}^{5/4}~~.
\end{equation}
   Thus, the efficiency of annihilation ranges from $\approx$0.1 to $
0.84 \times L_{53}^{5/4}$.  For the upper range of luminosity the
efficiency is quite large.  Also, using the supernova code of Wilson
and Mayle \cite{jwg:wm93} we calculate the entropy per baryon of the
plasma to be as high as $10^6$, thus the resulting pair plasma will
have low baryon loading.

\section*{Pair Plasma Expansion AND SHOCK WITH ISM}

\begin{figure}
\centerline{\epsfig{figure=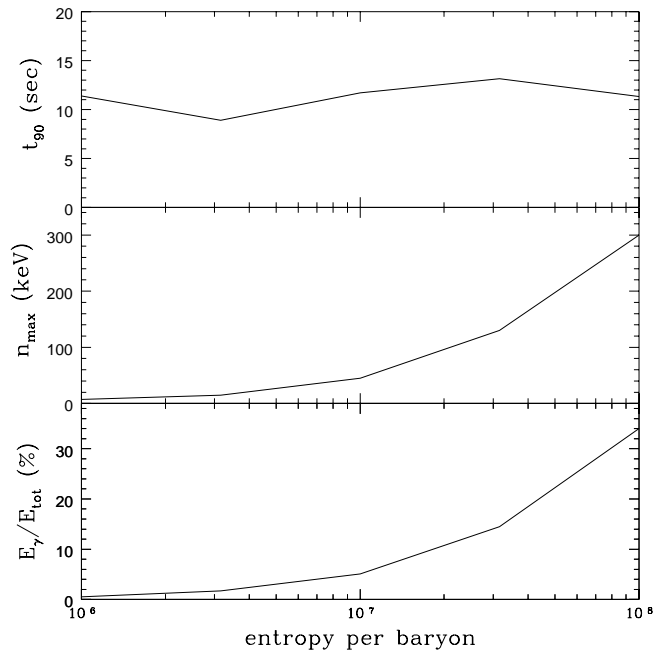, width=9cm}}
\caption{The duration, energy at the number spectrum peak, and
gamma-ray efficiency are plotted for the emission from an expanding
pair plasma with total deposited energy $E_{tot} = 10^{52}$ ergs over
a range of entropies per baryon $10^6$ to $10^8$.}
\label{jwg:date52}
\end{figure}

Having determined the initial conditions of the hot $e^+e^-$ pair
plasma near the surface of a neutron star, we wish to follow its
evolution and characterize the observable gamma-ray emission.  To
study this we have developed a spherically symmetric, general
relativistic hydrodynamic computer code to track the flow of baryons,
$e^+e^-$ pairs, and photons.  For the present discussion we consider
the plasma deposited at the surface of a $1.45 M_\odot$ neutron star
with a radius of 10 km.  Discussion of this code can be found in
\cite{jwg:swm97,jwg:swm00}.  In those papers the emission from an
expanding fireball was studied.  In Figure \ref{jwg:date52} it is
shown that the resulting emission spectrum and $\gamma$-ray emission
efficiency $E_\gamma/E_{tot}$ strongly depends upon the entropy per
baryon of the plasma deposited near the surface of the neutron stars;
entropies of $\lesssim 10^6$ resulted in weak emission with most of
the original energy manifesting itself as kinetic energy of the
baryons.  Thus, for the low entropy per baryon fireballs ($s \sim 10^5
- 10^6$) produced by NSBs it is necessary to examine the emission due
to the interaction of the relativistically expanding baryon wind with
the interstellar medium (ISM).  We find that these baryon winds
typically have a Lorentz factor $\gamma \approx 300$ and have a total
energy $\approx 10^{52}$ ergs.

After becoming optically thin and decoupling with the photons, the
matter component of the fireball continues to expand and interact with
the ISM via collisionless shocks.  As the ISM is swept up, the matter
decelerates.  We model this process as an inelastic collision between
the expanding fireball and the ISM as in, for example,
\cite{jwg:piran98}.  We assume that the absorbed internal energy is
immediately radiated away.  From this we construct a simple picture of
the emission due to the matter component of the fireball
``snowplowing'' into the ISM of baryon number density $n$.

\begin{figure}
\centerline{\epsfig{file=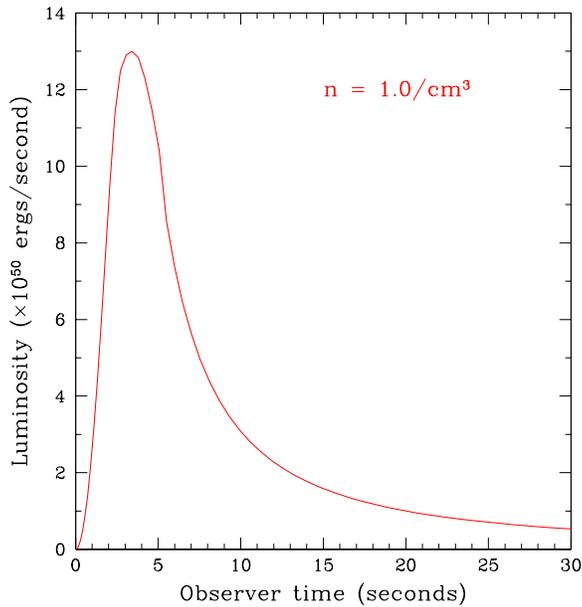, width=9cm}}
\caption{The light curve for a $10^{52}$ erg fireball expanding at
$\gamma = 300$ into the interstellar medium with number density
of one baryon per cm$^3$.}
\label{jwg:ltcurve}
\end{figure}

We have constructed an analytic formula for the luminosity in time
\cite{jwg:swm00} of the fireball plowing into the ISM.  We show a plot
of this function in Figure \ref{jwg:ltcurve} for a range of ISM
densities.  Defining $t_{max}$ as the time of maximum luminosity
\begin{equation}
L(t) \propto
	\begin{cases}
	t^2& \text{free expansion phase     $(t < t_{max})$} \\
	t^{-10/7}& \text{deceleration phase     $(t > t_{max})$.}
	\end{cases}
\end{equation}
This luminosity curve has the so called ``FRED'' (Fast Rise,
Exponential Decay) profile which is characteristic of real bursts.

\subsection*{Synchrotron Shock Spectrum}

\begin{figure}
\centerline{\epsfig{figure=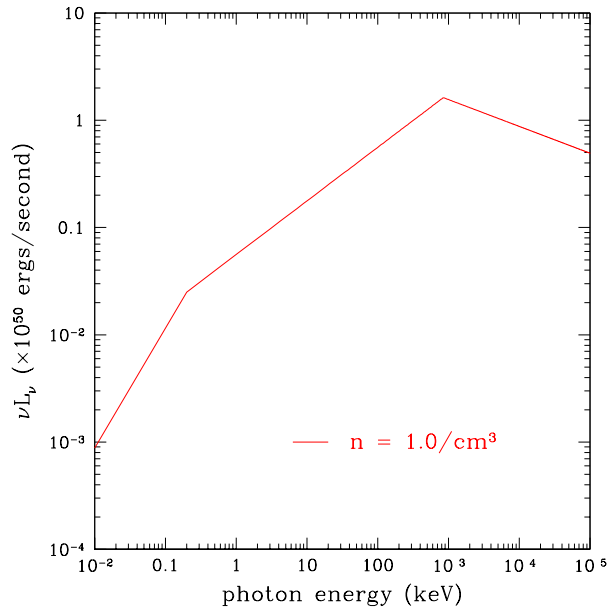, width=9cm}}
\caption{The peak synchrotron spectrum for a $10^{52}$ erg fireball
expanding at $\gamma = 300$ into interstellar medium number density of
one baryon per cm$^3$.}
\label{jwg:spec}
\end{figure}

Using the theory of synchrotron shocks (e.g. \cite{jwg:spn98}) we can
construct a spectrum as shown in Figure \ref{jwg:spec}.  To model the
synchrotron spectrum there are three free parameters: $\epsilon_B,
\epsilon_e$ are the fractions of baryonic kinetic energy that is
deposited into the magnetic field and the electrons respectively, and
$n$, the number density of baryons in the ISM.  In these calculations
we assume $\epsilon_B = \epsilon_e = 1/4$.  As shown in Figure
\ref{jwg:spec}, a reasonable ISM density of 1 baryon/cm$^3$ gives a
peak in the $\nu L_\nu$ spectrum at $\sim 100$ keV in agreement with
observations.  Calculations of the efficiency show that 75 \% of the
energy is emitted at photon energies of 10 keV and above
\cite{jwg:swm00}.

\section*{Conclusions}

In this proceedings we have argued that heated neutron stars (perhaps
by stellar compression of close neutron-star binaries) are viable
candidates for the production of large, high entropy per baryon,
$e^+e^-$ pair plasma fireballs, and thus, for the creation of
gamma-ray bursts.  We find that fireballs of total energies $E \sim
10^{51}$ to $3 \times 10^{52}$ ergs and entropies per baryon $s >
10^5$ are possible.  Also, this model gives a power-law spectrum that
peaks at hundreds of keV and has an overall efficiency of 10-20 \%.


Work performed under the auspices of the U.S. Department of Energy by
the Lawrence Livermore National Laboratory under contract
W-7405-ENG-48.  J.R.W. was partly supported by NSF grant PHY-9401636.
Work at University of Notre Dame supported in part by DOE grant
DE-FG02-95ER40934, NSF grant PHY-97-22086, and by NASA CGRO grant
NAG5-3818.  \\

\end{document}